\documentclass[usenatbib]{mnras}
\usepackage{graphicx, amssymb, aas_macros, natbib, bm}
\usepackage{ marvosym }
\usepackage{enumitem}
\usepackage{fancybox}
\usepackage{lipsum}  
\usepackage{gensymb}
\usepackage{csquotes}
\usepackage{xcolor}
\graphicspath{{plots/}}

\newcommand{\mhalo}{{M}_{\rm{halo}}}

\newcommand{\mstar}{{M}_{\star}}

\newcommand{\msun}{{\rm M}_{\odot}}

\newcommand{\lt}{<}
\newcommand{\gt}{>}

\DeclareRobustCommand{\ion}[2]{%
\relax\ifmmode
\ifx\testbx\f@series
{\mathbf{#1\,\mathsc{#2}}}\else
{\mathrm{#1\,\mathsc{#2}}}\fi
\else\textup{#1\,{\mdseries\textsc{#2}}}%
\fi}

\title[Studying Satellite Quenching with Machine Learning]
{A Machine Learning Approach to Measuring the Quenched Fraction of Low-Mass Satellites Beyond the Local Group}

\author[Baxter et al.]
{Devontae C. Baxter,$^1$\thanks{$\!\!$e-mail: dbaxter@uci.edu}\thanks{$\!\!$LSSTC DSFP Fellow}\ 
M. C. Cooper,$^1$\ 
Sean P. Fillingham$^2$\ 
\\
$\!\!^1$Center for Cosmology, Department of Physics \& Astronomy,
University of California, Irvine, 4129 Reines Hall, Irvine, CA 92697, USA \\
$\!\!^2$Department of Astronomy, University of Washington,
Box 351580, Seattle, WA 98195, USA \\
}

\begin{document}

\pagerange{\pageref{firstpage}--\pageref{lastpage}} 
\pubyear{2020}

\maketitle

\label{firstpage}
\begin{abstract}

Observations suggest that satellite quenching plays a major role in the build-up of passive, low-mass galaxies at late cosmic times. Studies of low-mass satellites, however, are limited by the ability to robustly characterize the local environment and star-formation activity of faint systems. 
In an effort to overcome the limitations of existing data sets, we utilize deep photometry in Stripe 82 of the Sloan Digital Sky Survey, in conjunction with a neural network classification scheme, to study the suppression of star formation in low-mass satellite galaxies in the local Universe. Using a statistically-driven approach, we are able to push beyond the limits of existing spectroscopic data sets, measuring the satellite quenched fraction down to satellite stellar masses of ${\sim}10^7~\msun$ in group environments ($\mhalo = 10^{13-14}~h^{-1}~\msun$). At high satellite stellar masses ($\gtrsim 10^{10}~\msun$), our analysis successfully reproduces existing measurements of the quenched fraction based on spectroscopic samples. Pushing to lower masses, we find that the fraction of passive satellites increases, potentially signaling a change in the dominant quenching mechanism at $\mstar \sim 10^{9}~\msun$. Similar to the results of previous studies of the Local Group, this increase in the quenched fraction at low satellite masses may correspond to an increase in the efficacy of ram-pressure stripping as a quenching mechanism in groups.

\end{abstract}

\begin{keywords}
  galaxies: evolution -- galaxies: star formation --
  galaxies: dwarf -- galaxies: formation 
\end{keywords}

\section{Introduction}
\label{sec:intro}

The recent generation of large-scale galaxy surveys have revealed that the population of non-star-forming (i.e.~\textquote{quiescent} or \textquote{quenched}) galaxies increased by more than a factor of two in the past $7 - 10$~Gyr, such that quenched systems, as opposed to their star-forming counterparts, comprise the majority of the stellar mass budget at $z \sim 0$ \citep{Bell04, Bundy06, Faber07}. While the growth of the global quenched population is relatively well measured at late cosmic time, our current understanding of the physical processes responsible for the suppression (or \textquote{quenching}) of star-formation remains woefully incomplete as evidenced by many models of galaxy formation overestimating the observed satellite quenched fraction \citep{Kimm09, Weinmann12, Hirschmann14}. Nevertheless, a wide assortment of physical processes have been put forth to explain how galaxies transition from star forming to quiescent. In general, these processes are split into two distinct categories, namely, internal and environmental quenching. The former, which acts independent of local environment (i.e.~on both central and satellite systems), refers to any quenching process that suppresses star formation from within a galaxy. Examples of internal quenching mechanisms include feedback from star formation \citep{Oppenheimer06, Ceverino09} and active galactic nuclei \citep[AGN,][]{DiMatteo05, Hopkins05, Croton06}. On the other hand, environmental quenching, which typically applies to low-mass satellites ($\lesssim 10^{10}~\msun$), refers to a range of quenching mechanisms that suppress star formation due to environmental factors -- e.g.~ram-pressure stripping \citep{GG72, Abadi99}, tidal stripping \citep{Merritt83, Moore99, Gnedin03}, strangulation or starvation \citep{Larson80, Kawata08}, and harassment \citep{Moore96, Moore98}. In general, environmental quenching mechanisms suppress star formation by either preventing satellites from accreting gas (e.g.~strangulation) or by removing pre-existing gas reservoirs through galaxy-galaxy interactions (i.e.~harassment), gravitational tidal forces (e.g.~tidal stripping), or interaction with the circumgalactic medium of the host (e.g.~ram-pressure stripping).

At $z \sim 0$, galaxy surveys find that satellites, not centrals, comprise the largest fraction of passive systems over a wide range of stellar masses ($\lesssim 10^{10.7}~\msun$, \citealt{Wetzel13}). Furthermore, observations of low-mass ($\lesssim 10^{9}~\msun$) galaxies in the local Universe have demonstrated that nearly all field galaxies are star forming, signaling that environmental quenching is primarily responsible for suppressing star formation in the low-mass regime \citep{Haines08, Geha12}. Altogether, these observations demonstrate the importance and ubiquity of satellite quenching at late times and especially low satellite masses. Yet, hydrodynamic and semi-analytic models, which successfully predict the fraction of quiescent centrals, continue to significantly overpredict the relative number of passive satellites, especially at low-masses (\citealt{Kimm09, Hirschmann14, Wang14}, but see also \citealt{Henriques17}). This discrepancy between theoretical predictions and observations is driven by a failure to properly model the physical processes responsible for satellite quenching. This lack of agreement between observations and theoretical models further emphasizes that understanding the details of satellite quenching is tantamount to advancing our understanding of galaxy formation.

Our current understanding of satellite quenching at low masses is largely derived from studies of dwarf galaxies ($\mstar \sim 10^{6}-10^{8}~\msun$) in the very local Universe, including our own Local Group. First and foremost, a range of observations demonstrate that the vast majority of low-mass satellites are gas-poor and passive, in contrast to their gas-rich, star-forming counterparts in the field \citep[e.g.][]{Grcevich09, Spekkens14, Weisz14a, Weisz14b}. 
Furthermore, studies of the accretion history of these systems using $N$-body simulations demonstrate that quenching is highly efficient, such that the typical timescale over which quenching occurs is $\sim2~{\rm Gyr}$ at $\mstar \lesssim10^{8}~\msun$ (likely driven by an increase in the efficacy of ram-pressure stripping, \citealt{Fillingham15, Fillingham16, Fillingham18, Wetzel15, Weisz15}). 
On the other hand, studies of the more massive satellites ($\mstar \gtrsim 10^{8}~\msun$) in the Local Group and nearby groups/clusters find that these systems have significantly longer quenching timescales ($\gtrsim 5~{\rm Gyr}$), consistent with starvation acting as the dominant quenching mechanism \citep{DeLucia12,Wetzel13,Wheeler14}. Taken together, this implies that a transition in the dominant quenching mechanism occurs at $\mstar \sim 10^{8}~\msun$ \citep[at least within Milky Way-like host halos, $\mhalo \sim 10^{12}~\msun$,][]{Fillingham16, RW19}. 

A major step towards increasing our understanding of satellite quenching involves determining if the aforementioned results extend beyond the Local Group. 
Is a similar increase in the quenched fraction observed at low masses outside of the Local Group (and/or in more massive host halos), indicating a corresponding increase in the efficiency of environmental (or satellite) quenching at this mass range? 
Unfortunately, the current generation of spectroscopic surveys lack the necessary combination of depth, area, and/or completeness to reliably probe this mass regime. 
For example, at the magnitude limit of the main spectroscopic survey, the Sloan Digital Sky Survey \citep[SDSS,][]{York00} can only probe galaxies with stellar masses less than $10^{8}~\msun$ at $z < 0.01$. 
While more recent surveys push fainter, including the Galaxy and Mass Assembly (GAMA) \citep[GAMA,][]{Driver09, Driver11} survey, the corresponding area of sky mapped is significantly smaller, again limiting the number of nearby hosts around which we can study their satellites.
In contrast to spectroscopic data sets, wide and deep imaging programs are able to probe both star-forming and passive galaxies down to stellar masses of $\sim10^{7}~\msun$ at $z \lesssim 0.1$, covering significant areas on the sky.

Herein, we present a method for measuring the satellite quenched fraction down to $\mstar \sim 10^{7}~\msun$ by applying machine learning and statistical background subtraction techniques to wide and deep photometric data sets, pushing beyond the limits of current spectroscopic samples. In \S\ref{sec:data}, we describe the spectroscopic and photometric data sets utilized in our analysis. In \S\ref{sec:NNC}, we discuss the training, testing, and performance of our neural network classifier (NNC) as well as our use of the trained model to classify galaxies in our photometric sample as star forming or quenched. In \S\ref{sec:backsub}, we describe our statistical background subtraction technique and use it to measure the satellite quenched fraction around nearby groups. Lastly, in \S\ref{sec:summmary_and_discussion}, we discuss and summarize our results. When necessary, we adopt a flat $\Lambda$CDM cosmology with $H_{0} = 70~{\rm km}~{\rm s}^{-1}~{\rm Mpc}^{-1}$ and $\Omega_{m}$ = 0.3. All magnitudes are on the AB system \citep{OkeGunn83}.


\section{Data}
\label{sec:data}

\subsection{Photometric Sample}
\label{subsec:S82}

Our analysis utilizes the co-added images and photometry from the Sloan Digital Sky Survey, 
focusing on the deeper Stripe 82 data set \citep[S82,][]{Annis14, Bundy15}. Stripe 82 is centered on the Celestial Equator and is comprised of an area of $\sim~300~{\rm deg}^{2}$ that spans between $-50 \degree \lt \alpha \lt 60 \degree$ and $-1.25 \degree \lt \delta \lt +1.25 \degree$. The co-added images in S82 reach a depth $\sim2$ magnitudes deeper in $ugriz$ relative to the SDSS single-pass data. Overall, the wide area and impressive depth ($r \sim 22.4$, 95\% complete for galaxies) of Stripe 82 make it well-suited for studying the properties of low-mass galaxies in the local Universe. For the purpose of our analysis, we limit the S82 sample to only include galaxies (defined using the SDSS \texttt{TYPE} parameter) with $13 \lt r \lt 21.5$. This apparent $r$-band magnitude cut is applied to ensure that galaxies in our sample are below the SDSS saturation limit and above the $95\%$ completeness limit for galaxies in the $gri$ passbands.

We exclude the shallower and less complete $u$ and $z$ bands throughout our analysis. 
Furthermore, as discussed in \cite{Bundy15}, the \texttt{TYPE}-based galaxy classification is contaminated with a non-negligible fraction of stars ($\sim 10\%$), which is attributed to PSF characterization issues in the co-added images. Based on visual inspections, we find that the fraction of stars misclassified as galaxies is higher and more pronounced at brighter magnitudes. 
To eliminate stars from the photometric sample, we remove sources at $15 < r < 18$ that are classified as stars in the corresponding single-pass SDSS images. At the very brightest magnitudes ($r < 15$), where number counts are lower, we remove stars based on a visual inspection of the single-pass SDSS images. Combined, these two procedures remove $\sim 6\%$ of sources at $r \lt 18$ from our sample, such that our final catalog includes $1{,}293{,}392$ galaxies with non-extinction-corrected $gri$ photometry in S82.
Accounting for Galactic extinction does not change our qualitative results, in part due to the low extinction in the S82 field \citep{SFD98}.

\subsection{Spectroscopic Training Set}
\label{subsec:MPA}

To train our classification scheme, which aims to identify galaxies as star forming or quenched, we use spectroscopic data products from the Max Planck Institute for Astrophysics and Johns Hopkins University DR7 catalog \citep[MPA-JHU,][]{Kauffmann03, Brinchmann04} along with photometry from SDSS Data Release 7 \citep{Hiroaki2011}. The MPA-JHU catalog is a value-added data set derived from the spectroscopic SDSS DR7, containing stellar mass and star formation rate (SFR) estimates for nearly a million 
galaxies up to $z \sim 0.3$. When available, the SFRs are derived using the extinction-corrected H${\alpha}$ emission line luminosities. For galaxies that lack emission lines, the SFRs are estimated using a relationship between SFR and the 4000\AA-break index \citep[$D_{4000}$,][]{Bruzual83, Hamilton85, Brinchmann04}. Likewise, the stellar masses are computed using model fits to the broad-band $ugriz$ photometry \citep{Kauffmann03}. 

We match galaxies in the MPA-JHU and SDSS DR7 catalogs using their unique MJD, plate ID, and fiber ID to construct a cross-matched catalog that includes both photometric and spectroscopic galaxy properties. These properties include $gri$ model magnitudes, specific star-formation rates (sSFR; SFR divided by stellar mass), redshifts, and stellar masses. 
Furthermore, we limit our cross-matched catalog to only include galaxies in which CLEAN = 1, RELIABLE $\neq$ 0. The former is a photometric flag that removes sources suffering from saturation, deblending, and/or interpolation issues. The latter is a spectroscopic flag that omits galaxies with unreliable line profiles and physical parameters. Overall, these cuts remove roughly $3\%$ of galaxies from the original MPA-JHU catalog. 
Finally, we limit our sample to only include galaxies at $z < 0.1$ and $\mstar \gt 10^{6.5}~\msun$, with measured specific star formation rates. Overall, our final sample includes $\sim 240{,}000$ galaxies, with a median redshift of $0.07$, median stellar mass of $2.7 \times 10^{10}~\msun$, and median $r$-band magnitude of $17$. 

\subsection{Host Sample}
\label{subsec:Yang}

Our spectroscopically-confirmed host sample is selected from the group catalog of \citet{Yang07}. We select groups within the Stripe 82 footprint at $z \lt 0.1$ and $10^{13} < \mhalo~h^{-1}~\msun < 10^{14}$, excluding groups that are located within $0.5$~degrees of the edges of the Stripe~82 field. Our final sample consists of $110$ hosts, with a median redshift of $0.077$ and a median halo mass of $1.6 \times 10^{13}~\msun$. The central galaxies in these groups have a median stellar mass of $1.3 \times 10^{11}~\msun$. 
%


\begin{figure}
 \centering
 \hspace*{-0.25in}
 \includegraphics[width=3.25 in]{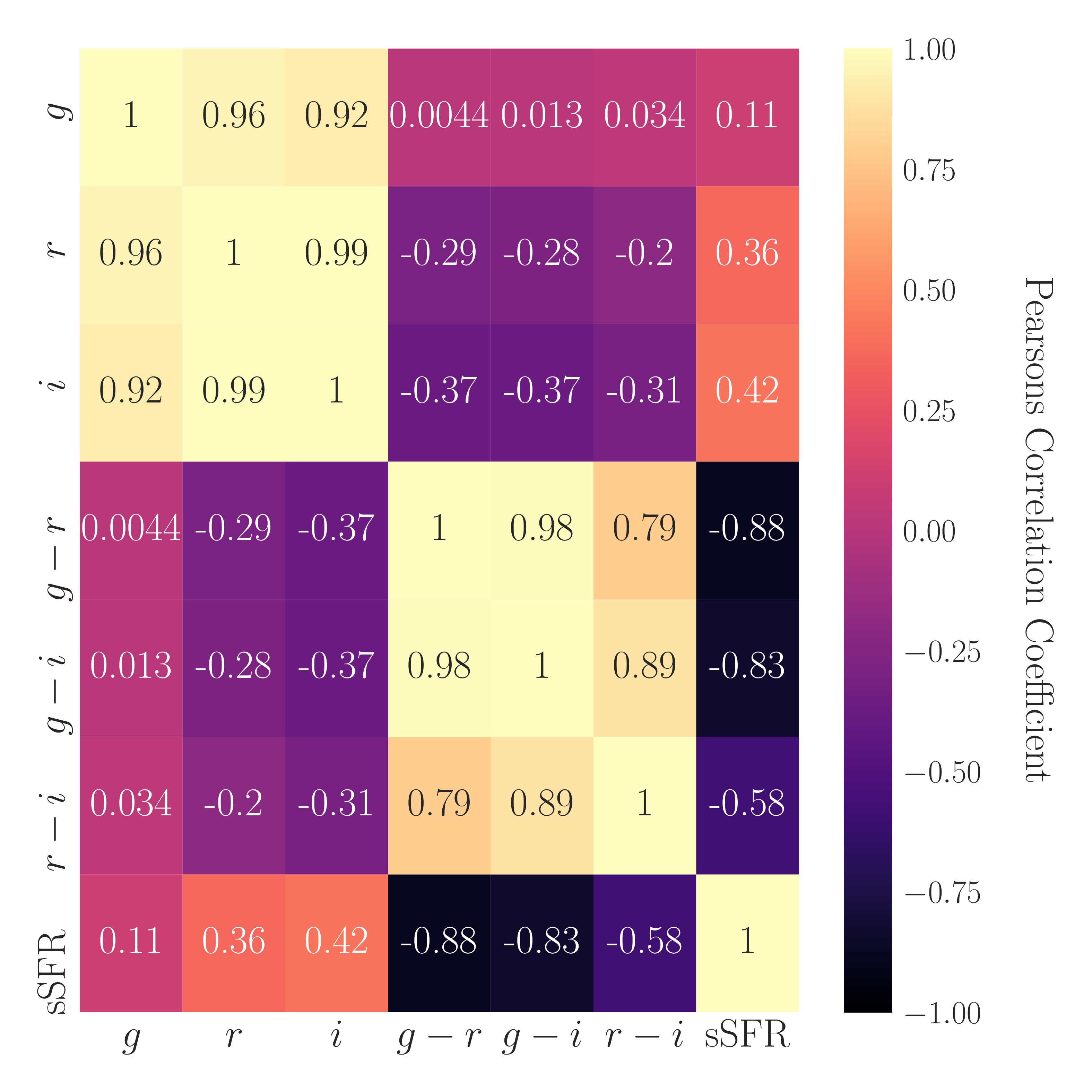}
 \caption{A heatmap displaying the correlation between observed colors, apparent magnitudes, and specific star-formation rates for galaxies in our spectroscopic training set. 
 In general, supervised neural network classifiers rely heavily on an existing correlation between input features and target variables (i.e.~quenched or star-forming label). For our sample, we detect a relatively strong correlation between observed color and sSFR.}
 \label{fig:CM}
\end{figure}


\section{Neural Network Classifier}
\label{sec:NNC}

\subsection{Feature Selection and Pre-Processing}
\label{subsec:3.1}
The first step in constructing our training set for supervised machine learning involves selecting the appropriate features that will enable our machine learning model to accurately classify galaxies as either star forming or quenched. Moreover, we can only include photometric features since we ultimately seek to apply our neural network classifier (NNC) to galaxies without spectra. To that end, we construct a heatmap to visualize the degree of correlation, as measured by the Pearson correlation coefficient, between the specific star formation rate of the MPA-JHU galaxies and their photometric properties. 
As shown in Figure~\ref{fig:CM}, we find a relatively strong negative correlation between the optical colors of the galaxies and their specific star formation rates, which implies that optically blue (red) galaxies tend to have higher (lower) specific star formation rates. With this correlation in mind, we construct our training set using only the $g-r$, $r-i$, and $g-i$ observed colors as features. 
The inclusion of magnitude information (i.e.~apparent $gri$ magnitudes) has a negligible effect on the resulting classifications, and as such was not utilized in the final configuration. 

\begin{figure}
 \vspace*{0.225in}
 \includegraphics[width=3.20in]{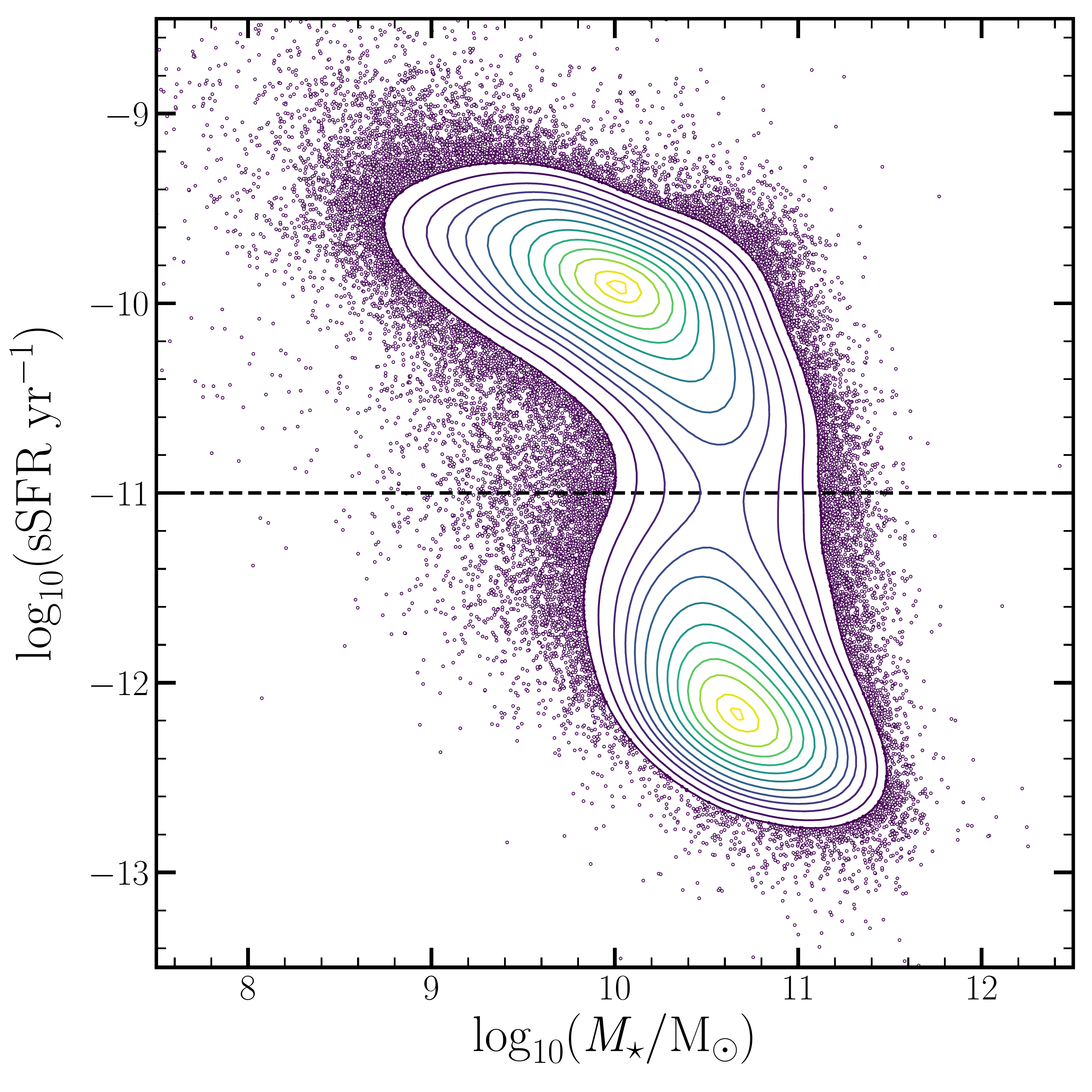}
 \caption{Specific star formation rate versus stellar mass for galaxies in our spectroscopic training set. The contours highlight the star-forming and quenched galaxy population within our sample. We divide the galaxy sample at ${\rm sSFR} = 10^{-11}~{\rm yr}^{-1}$, such that galaxies above this threshold are labeled as star forming and galaxies below this threshold are labeled as quenched.}
 \label{fig:MPA-JHU}
\end{figure}


\begin{figure*}
 \centering
 \hspace*{-0.1in}
 \includegraphics[width=6.0in]{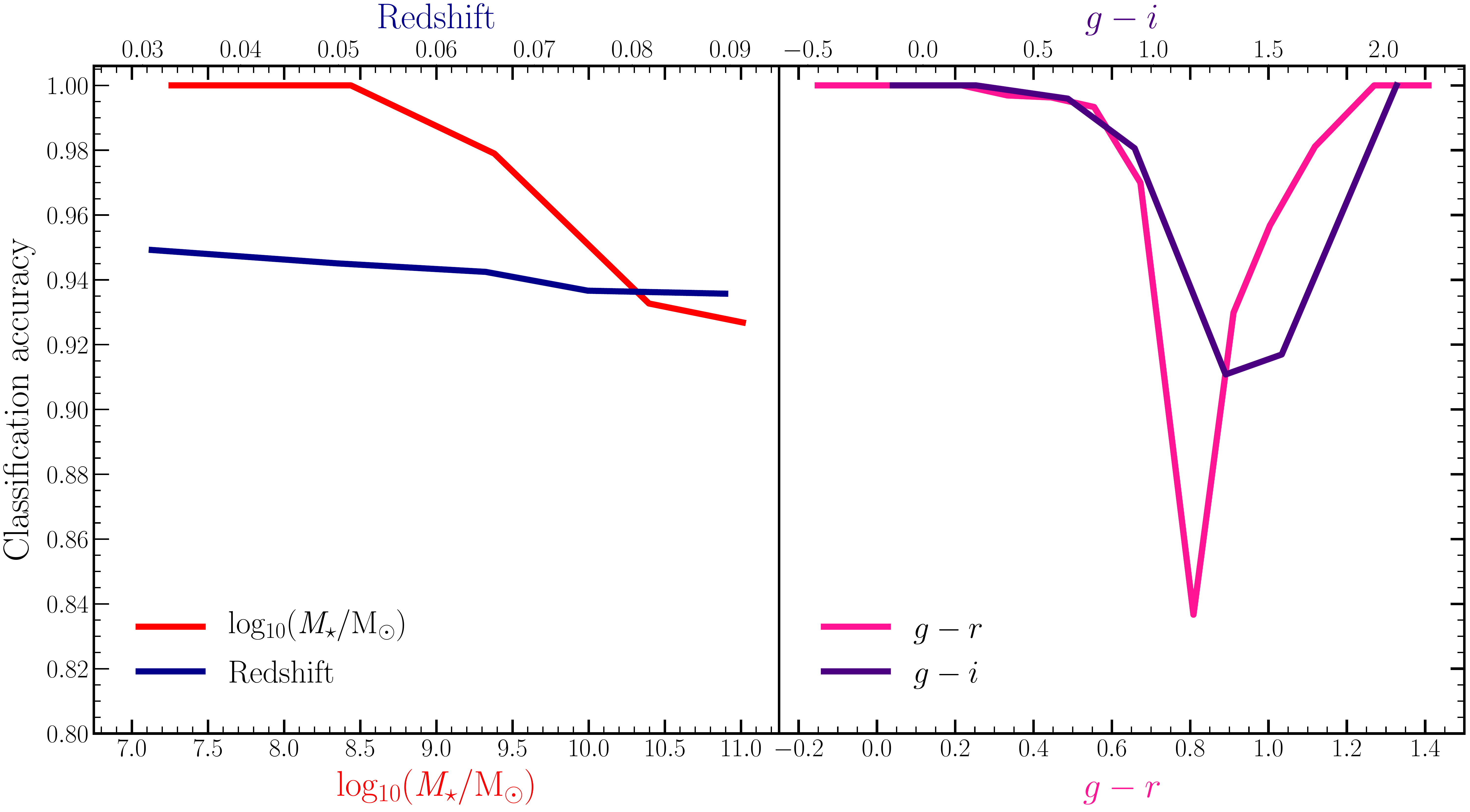}
 \caption{
 Classification accuracy in the validation set as a function of redshift and stellar mass (\emph{left}) along with observed $g-r$ and $g-i$ color (\emph{right}). The accuracy of the NNC is largely independent of host halo redshift and weakly dependent on stellar mass, with high-mass galaxies more likely to be incorrectly classified. Despite the training set being largely composed of high-mass galaxies ($\sim90\%$ of the training set has $\mstar >10^{9.5}~\msun$), we find that the overall classification accuracy as well as our primary satellite quenched results remain qualitatively unchanged when high-mass systems are omitted from the training set. As expected, the NNC is less reliable at classifying galaxies at intermediate colors (i.e.~in the \textquote{green valley} of the color bimodality) precisely due to the binary nature of the classification scheme.} 
 \label{fig:ca}
\end{figure*}


The second important step in constructing our training set for supervised machine learning involves systematically labeling galaxies as either star forming or quenched. We achieve this by taking advantage of the strong bimodality in sSFR-$\mstar$ space, which for our MPA-JHU sample is illustrated if Figure~\ref{fig:MPA-JHU}. In particular, we adopt a cut of ${\rm sSFR} = 10^{-11}~{\rm yr}^{-1}$ as our quenching threshold, such that galaxies above (below) this threshold are labeled as star forming (quenched). This results in a balanced training set where 49\% (51\%) of galaxies are classified as quenched (star forming). This is important because imbalanced training sets can result in uninformative models that naively overpredict the majority class and underpredict the minority class. Furthermore, we standardize the features of our training set to have a mean of zero and standard deviation of one according to $X_{\rm st} = (X-\mu)/\sigma$, where $X$, $\mu$, and $\sigma$ are the input feature, mean, and standard deviation of the sample, respectively. This pre-processing procedure is implemented to optimize the performance and stability of the neural network classifier, which assumes that the inputs are standardized.

Lastly, to construct our validation set, we remove $6600$ out of the $\sim 240{,}000$ galaxies in our training set. The validation set is composed of a subset of those galaxies cross-matched between the MPA-JHU and the S82 photometric catalogs using a search radius of 1$\arcsec$. We omit these galaxies from the training and testing process, so that they can ultimately be used to evaluate the performance of the resultant NNC. 
%


\subsection{Supervised Neural Network Classifier}
\label{subsec:3.2}

The supervised neural network classifier is a machine learning model that is trained using labeled observations in order to learn a mapping function between input features and output targets. The utility of these models is that once they are trained they can be readily used to classify unlabeled observations. Moreover, neural network classifiers are constructed using a variety of hyperparameters that influence the overall performance of the machine learning model. The optimal hyperparameters for our NNC are obtained using a K-fold cross-validation grid search. The names and values of these hyperparameters are as follows: (i) the number of hidden layers is two; (ii) the number of nodes in the first and second hidden layer are 8 and 4, respectively; (iii) the batch size is 64; (iv) the number of epochs is 10; (v) the dropout is 20\%. As is standard for binary classification, we use the rectified linear unit (ReLU) activation function for the input and hidden layers, while the sigmoid activation function is used for the output layer. Our model is compiled using a binary cross-entropy loss function and stochastic gradient descent with a learning rate of 0.01. Lastly, we use a stratified K-fold cross validation procedure with k=5 to determine the average accuracy and logarithmic loss of our model.

\subsection{Performance of Neural Network Classifier}
\label{subsec:3.3}

The K-fold cross validation yields an average classification accuracy of 0.94 and logarithmic loss of 0.17. Here, the accuracy measures the fraction of galaxies that are correctly classified during the training/testing process, while the logarithmic loss measures the uncertainty of the predictions made by the NNC. Therefore, the high average classification accuracy and low logarithmic loss suggest that our NNC returns both accurate and precise classifications. 
Another diagnostic for determining the reliability of the NNC involves applying the trained model to labeled data that was not utilized during the training or testing process. In our case, we use our validation set that is composed of a subset of the galaxies cross-matched between our spectroscopic training set and the S82 photometric sample. Upon applying the NNC to our validation set, we find that $93\%$ of galaxies in the validation set are correctly classified as quenched, and $95\%$ of star forming galaxies in the validation set are correctly classified as star forming. Moreover, we find that the true quenched fraction for the validation set is reproduced by the NNC with an average percent error of $\sim2\%$, largely independent of stellar mass.

In addition to classifying the galaxies in the validation set, the NNC also provides a classification probability (CP) between 0 and 1 for each prediction such that the CP equals 0 (1) when the model is 100 percent certain that a given galaxy is quenched (star forming). With this information, we define the classification confidence to be equal to the classification probability when ${\rm CP} \gt 0.5$ and equal to $1 - {\rm CP}$ when ${\rm CP} \lt 0.5$. The mean and median classification confidence are $0.927$ and $0.98$, respectively. 
Overall, these results provide further confidence in the reliability and accuracy of the predictions made by the neural network classifier.

Using the validation set, we also explore how the classification accuracy varies with galaxy properties. As shown in Figure~\ref{fig:ca}, we find that the classification accuracy remains relatively constant across our specified redshift range, such that host halos at slightly lower redshift are not biased relative to their higher-$z$ counterparts within the sample.  
We do find, however, a modest correlation between classification accuracy and stellar mass, such that the NNC achieves higher levels of accuracy when classifying lower-mass galaxies ($\mstar \lesssim 10^{9}~\msun$).   
While the spectroscopic training set is dominated by more massive galaxies ($\sim90\%$ of the spectroscopic training set has $\mstar >10^{9.5}~\msun$), the classification accuracy -- and our primary results regarding the satellite quenched fraction -- are qualitatively unchanged when limiting the spectroscopic training set to systems with $10^{6.5}~\msun \lt \mstar \lt 10^{9.5}~\msun$.

As suggested in Figure~\ref{fig:CM}, $g-r$ and $g-i$ color are the most informative features with respect to predicting whether a galaxy is star forming or quenched. In Figure~\ref{fig:ca}, we explore the relationship between the classification accuracy and these two features. As expected, the classification accuracy is highest for very blue and red galaxies, with a modest decrease for galaxies residing in the green valley. This is in part due to the binary nature of our classification scheme (i.e.~the lack of a transitory phase between star forming and quenched) along with the overlap between dusty star-forming galaxies and quiescent systems in rest-frame optical color \citep[e.g.][]{Yan06, Maller09, Williams09}. 
%


\begin{figure*}
 \centering
 \hspace*{-0.001in}
 \includegraphics[width=0.98\textwidth]{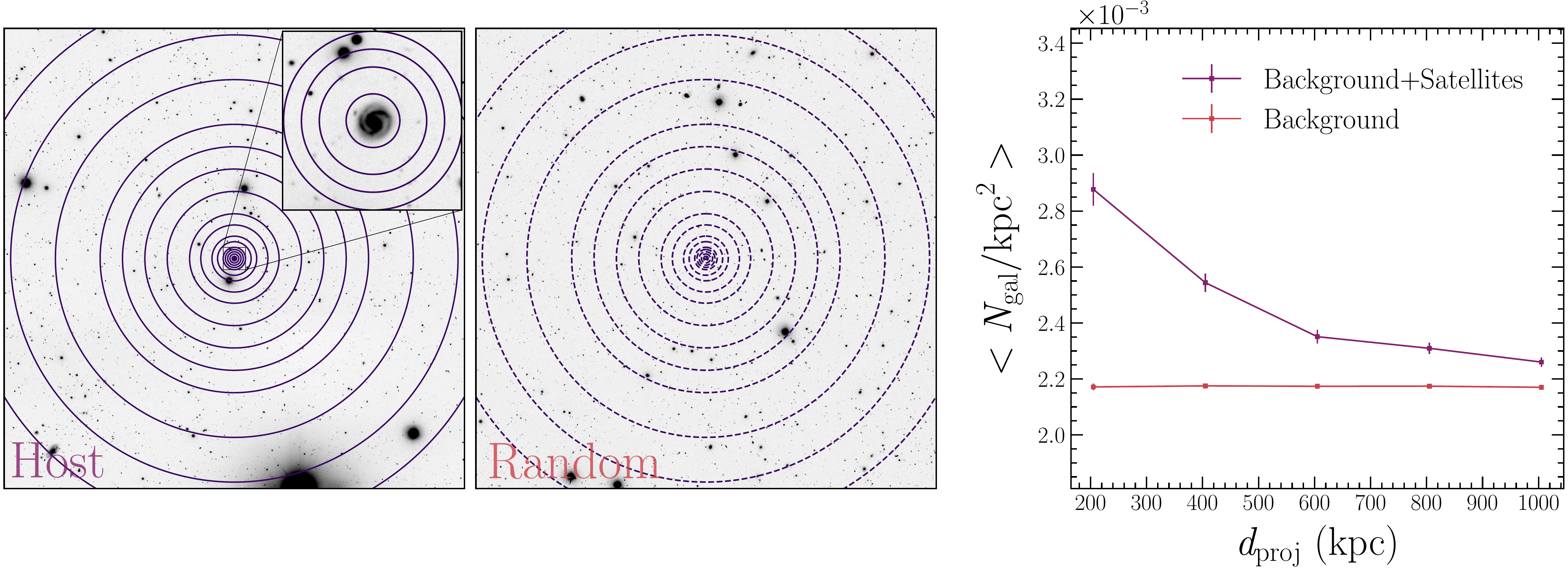}
 \caption{An illustration of our background subtraction technique, in which we measure the radial number density of galaxies around our host galaxies (\emph{left}) and in randomly-selected background fields (\emph{middle}). Using the photometric sample, we compute the mean number density profile as a function of projected radial distance, averaged over our sample of hosts and $6 \times 10^{4}$ background fields (\emph{right}). The error bars for both radial profiles correspond to 1-$\sigma$ Poisson errors in the measured surface density of galaxies.}
 \label{fig:backsub}
\end{figure*}


\subsection{Classification of Galaxies in Stripe 82}
\label{subsec:3.5}

Using the hyperparameters discussed in \S\ref{subsec:3.1}, we train our NNC on the entirety of the spectroscopic training set (\S\ref{subsec:MPA}). Next, we standardize the apparent $g-r$, $g-i$, and $r-i$ colors of the galaxies in our S82 photometric sample to have a mean of zero and standard deviation of one using the same procedure outlined in \S\ref{subsec:3.1}. Classifying our photometric sample using the trained NNC, we find that 66$\%$ of the galaxies in our photometric sample are classified as quenched while 34$\%$ are classified as star forming. We recognize that the fraction of passive galaxies in our photometric sample is biased high due to the inevitable inclusion of high-$z$ galaxies. Many of these high-$z$ sources have red apparent colors, and are more likely to be classified as quenched. Ultimately, the success of our approach relies on correctly classifying the low-$z$ sources (i.e.~the satellites of our targeted group sample). With that objective, the next step in our analysis involves combining the classification results with a statistical background subtraction technique to ultimately determine the satellite quenched fraction of our low-$z$ host sample. 
  

\section{Analysis of the S82 Sample}
\label{sec:backsub}

\subsection{Statistical Background Subtraction}
\label{subsec:bs}

While deep imaging allows the satellite population around nearby hosts to be detected and our NNC is able to robustly classify sources as star forming or quenched, identifying the true satellites amongst the sea of background sources remains a challenge. This is primarily due to the lack of highly-complete line-of-sight velocity information for our photometric sample, which is required to cleanly determine if a particular source is truly a satellite of a given host. However, instead of identifying properties of individual satellites, we employ a statistically-driven background subtraction technique that enables us to robustly measure the average properties of the satellite population. Figure~\ref{fig:backsub} illustrates our methodology, by which we compare the radial distribution of galaxies around nearby hosts to that measured in random positions on the sky. By subtracting the random background, we are able to measure the average properties (e.g.~radial profile, rest-frame color distribution, quenched fraction) of the underlying satellite population.

This statistical approach has proven effective in previous studies of satellites at intermediate redshift \citep{tal13, kaw14, nierenberg11, nierenberg12}. In general, the background subtraction procedure utilized in these studies involves measuring the radial distribution of galaxies around spectroscopically-confirmed hosts and subtracting the contribution from the background/foreground galaxies. For our analysis, we utilize the $110$ centrals from the \citet{Yang07} group catalog that overlap with the S82 footprint as our sample of spectroscopically-confirmed host galaxies. As stated in \S\ref{subsec:Yang}, our host sample is situated at $z \lt 0.1$ and have halo masses between $10^{13}$ and $10^{14}~h^{-1}~\msun$. 

Our technique for estimating the contribution from background galaxies involves measuring the radial distribution of galaxies at random positions within the S82 footprint. In particular, we generate $10^{6}$ random positions within Stripe 82, assigning each a corresponding redshift between $0.02 \lt z \lt 0.1$ as randomly drawn from a uniform distribution. As was done for the host sample, the random positions are also required to be less than $0.5$~degrees from the edges of the S82 field. We have also considered requiring the random points to be sufficiently far away from the spectroscopic hosts (e.g.~$d_{\rm proj} \gt 1-2~{\rm Mpc}$). However, we ultimately omitted this constraint since both scenarios return qualitatively similar results. 

We partition our hosts and random positions into six evenly-spaced redshift bins between $0.02 \lt z \lt 0.1$. For a given redshift bin, we count the number of quenched and star-forming galaxies in annuli centered on the hosts in bins of $r$-band magnitude. This procedure is repeated at the location of the background pointings, for which we count the number of quenched and star-forming galaxies in bins of $r$-band magnitude within annuli centered on $100$ random positions. Specifically, the photometric sample is partitioned into seven $r$-band magnitude bins between $13 \lt r \lt 21.5$ and the galaxies are counted in five annuli between $15$ and $1000$~kpc. For each $r$-band magnitude and redshift combination, we calculate the average number of quenched and star-forming galaxies per annuli for both the background and the spectroscopically-confirmed centrals. Moreover, for each individual host/random position, we calculate the 1-$\sigma$ Poisson error associated with our measurement and propagate this error in the calculation of the average number of galaxies per annuli. Increasing the number of random pointings used to determine the background (i.e.~$>100$) yields no significant change in our results.

\begin{figure}
 \centering
 \hspace*{-0.1in}
 \includegraphics[width=3.25in]{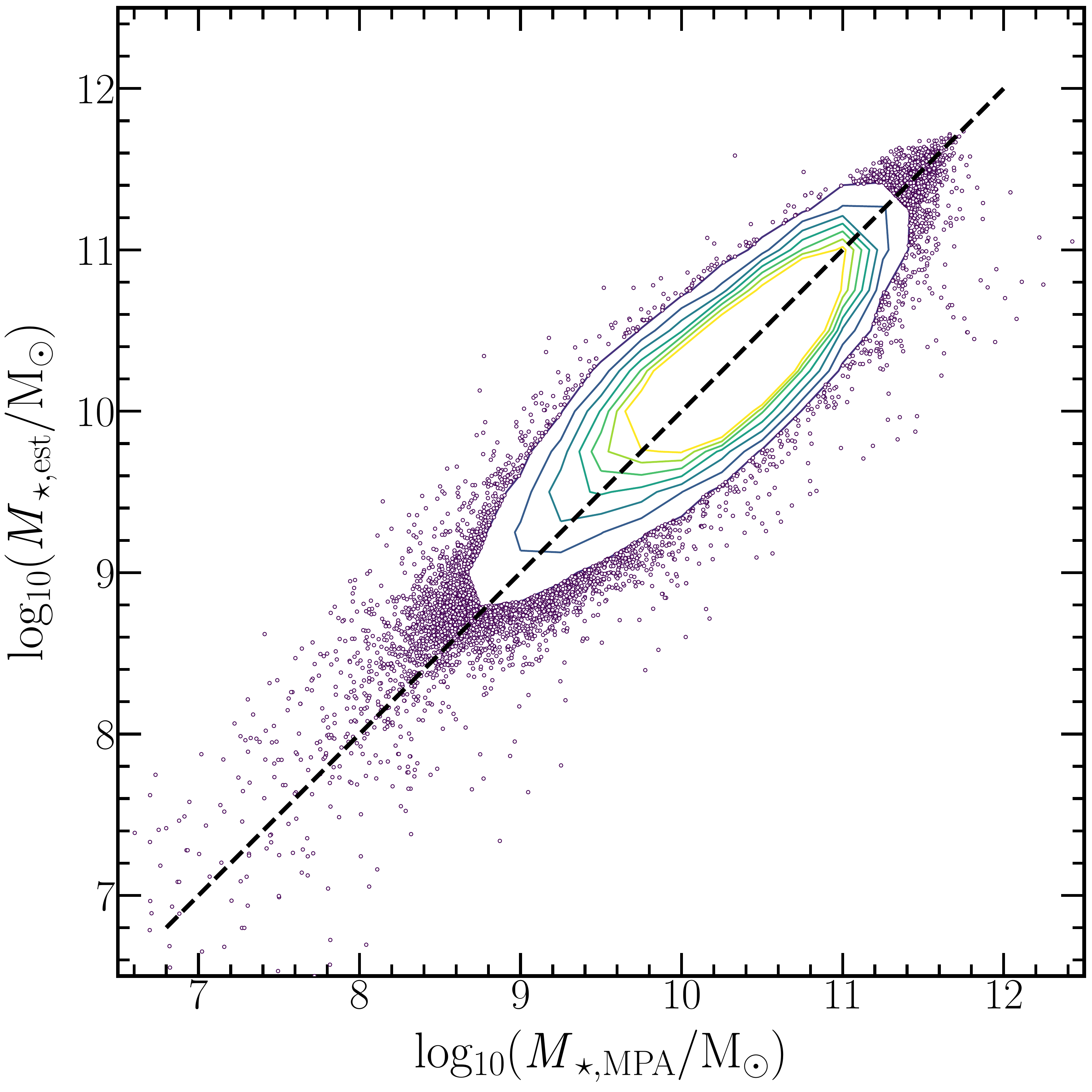}
 \caption{Comparison between our estimated stellar masses and those provided by the MPA-JHU catalog. Our stellar mass estimator, which we infer by fitting galaxies in the MPA-JHU catalog using Equation~\ref{equation:linear_reg}, provides robust mass measurements in the absence of multi-band photometry. In comparison to the MPA-JHU measurements, the median stellar mass difference is $-0.030$ dex with a standard deviation of $0.22$ dex. }
 \label{fig:smvr}
\end{figure}

The galaxies are counted in the manner outlined above because it allows us to robustly estimate stellar masses for our statistical satellite population by capitalizing on the strong correlation between apparent $r$-band magnitude and stellar mass at fixed redshift. To determine this mapping from $r$ and $z$ to stellar mass, we fit the following relation to galaxies in the MPA-JHU catalog 
\begin{equation}
    \mstar(r, z) = \gamma*r + b(z) \; ,
    \label{equation:linear_reg}
\end{equation}
where $\gamma$ and $b(z)$ correspond to the slope and $y$-intercept of the fit in a given redshift bin. In particular, we fit this relation in redshift bins (with typical width of $\Delta z = 0.005$), such that 
the statistically-inferred galaxy counts as a function of $r$-band magnitude (following background subtraction) can be mapped to stellar mass based on the redshift of the host system. 
In Figure~\ref{fig:smvr}, we compare the stellar masses estimated using our best-fit parameters for Equation~\ref{equation:linear_reg} 
to the corresponding stellar masses from the MPA-JHU catalog, which are based on fitting the multi-band photometry to model spectral energy distributions. 
Our stellar mass estimates, inferred solely from the observed $r$-band magnitude, are relatively accurate with a median difference of $-0.030$ dex and a $1\sigma$ scatter of $0.22$ dex. 
There is a slight bias towards our method under- and over-predicting the masses of high-mass and low-mass galaxies, respectively. 
Not surprisingly, the fits to Equation~\ref{equation:linear_reg} are best at intermediate stellar masses, where the spectroscopic training set is more abundant. 
Tuning our fits to better reproduce the stellar masses of low-mass systems does not yield a significant change in our results, with our measurements of the satellite quenched fraction computed in bins of stellar mass that exceed the typical measurement uncertainty.

Altogether, the statistical background procedure provides us with a measure of the average number of quenched and star-forming galaxies as a function of projected distance and stellar mass at the location of both the spectroscopically-confirmed host galaxies and the random background positions. With these galaxy counts and classifications, we compute the average number of quenched and star-forming satellites as a function of stellar mass and projected distance according to 
\begin{equation}
\bar{N}_{\rm sats}(d_{\rm proj}, \mstar) =\sum ( \bar{N}_{\rm back+sats} - \bar{N}_{\rm back})  \; ,
\label{equation:nsats}
\end{equation}
where $\bar{N}_{\rm back+sats}$ and $\bar{N}_{\rm back}$ are the average number of galaxies measured in annuli centered on the spectroscopically-confirmed centrals and random positions, respectively.


\begin{figure}
 \hspace*{-0.1in}
 \includegraphics[width=0.48\textwidth]{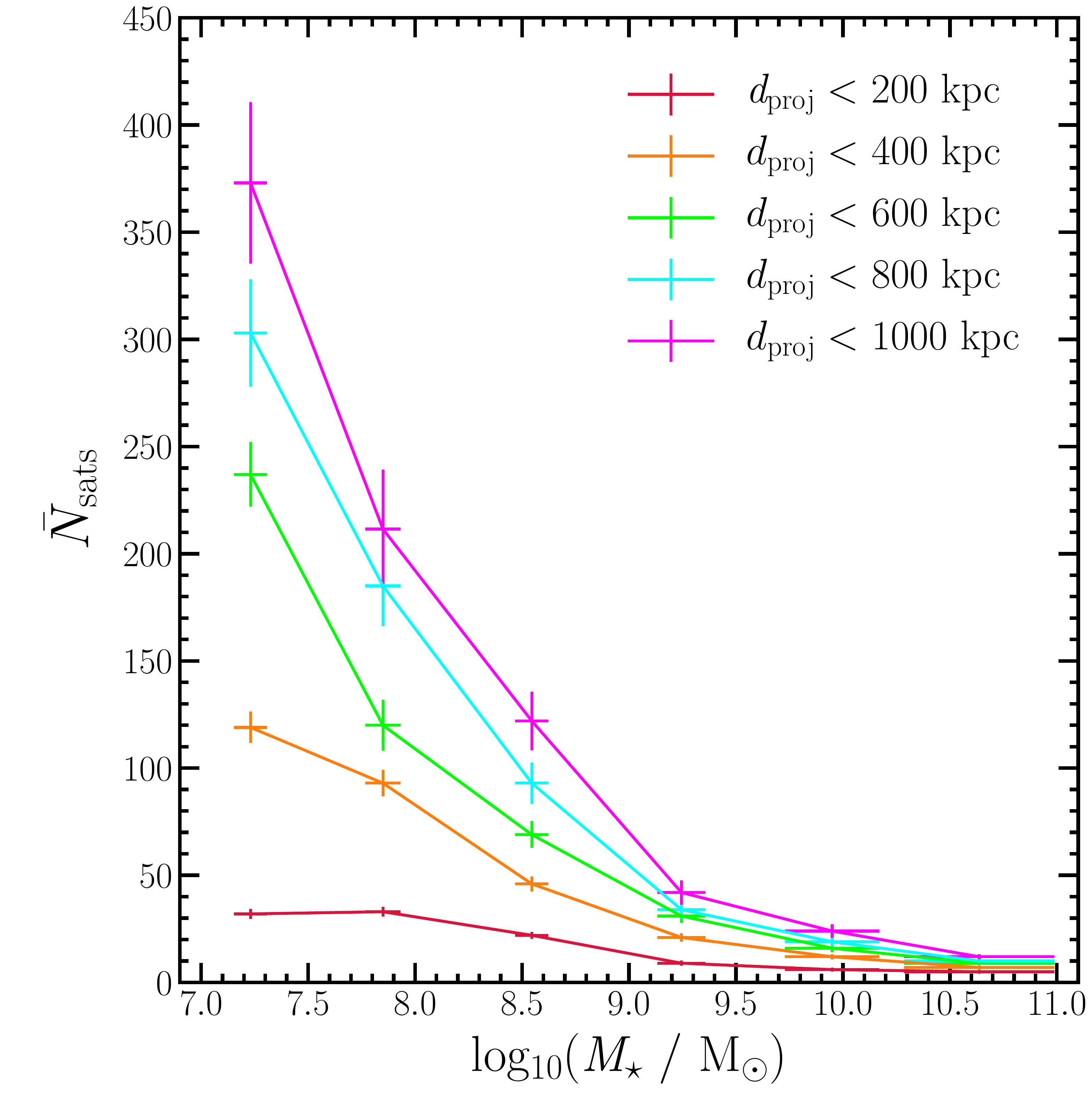}
 \caption{The average number of satellites as a function of stellar mass in projected distance bins. The vertical error bars gives the standard deviation in the distribution of the number of satellites after repeating the background subtraction procedure $100$ times, whereas the horizontal error bars represent the standard deviation within the stellar mass bin. For our analysis, we limit our satellite population to systems at $d_{\rm proj} \lt 400~{\rm kpc}$.}
 \label{fig:lum_func}
\end{figure}


In Figure~\ref{fig:lum_func}, we show the resulting average number of satellites as a function of stellar mass and projected host-centric distance. 
We adopt $400~{\rm kpc}$ as the outer extent of our groups (roughly $R_{200}$) based upon a comparison to similar halos in the IllustrisTNG project \citep{Nelson18, Nelson19a, Naiman18, Marinacci18, Springel18, Pillepich18}. 
For host halos at $z = 0$ and $10^{13}~h^{-1}~\msun< M_{200} < 10^{14}~h^{-1}~\msun$ within the TNG300 simulation, a sample of $>2000$ halos with a median mass of $M_{200} \sim 1.85 \times 10^{13}~h^{-1}~\msun$, the 
median $R_{200}$ is $430~h^{-1}~{\rm kpc}$ with a $1\sigma$ scatter of $97~h^{-1}~{\rm kpc}$. 
Given that our measurements are made in projection, we limit our selection of the satellite population to projected distances of $< 400~{\rm kpc}$. While this excludes a subset of satellites at host-centric distances of $400~{\rm kpc} < R < R_{200}$, it also reduces contamination from objects in the surrounding infall regions ($R \sim 1\mbox{-}2~R_{200}$).  
As discussed in \S\ref{subsec:qf}, our results are qualitatively unchanged when including satellites out to projected distances of $600~{\rm kpc}$ or $800~{\rm kpc}$.

\begin{figure}
 \hspace*{-0.1in}
 \includegraphics[width=0.48\textwidth]{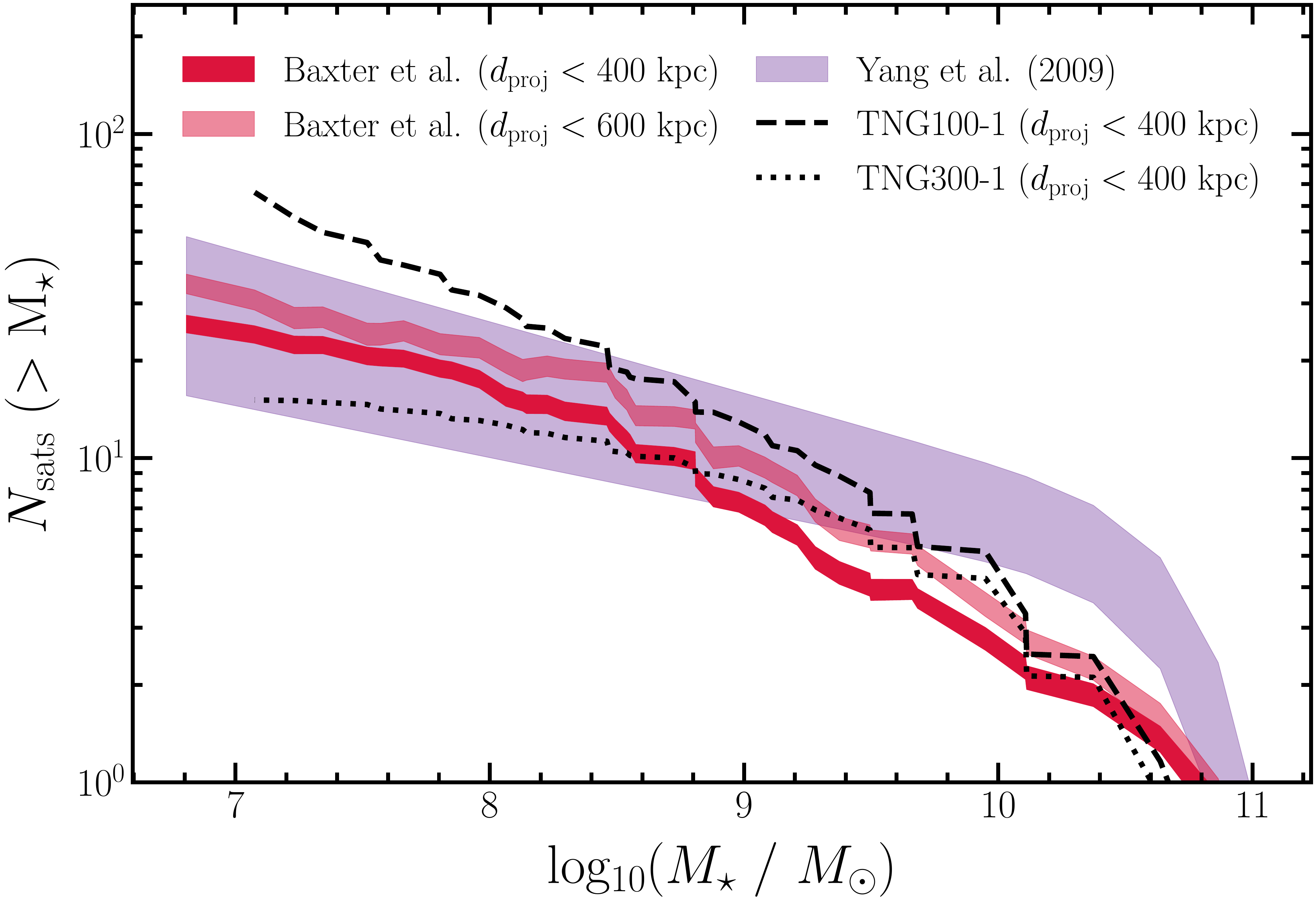}
 \caption{The cumulative satellite stellar mass function based on our statistical background subtraction technique in comparison to that from spectroscopic observations and simulations. The dark and light crimson bands show our satellite counts (per group) at $R < 400~{\rm kpc}$ and $R < 600~{\rm kpc}$, respectively. The purple band corresponds to the satellite stellar mass function for groups with $\mhalo = 10^{13.2-13.8}~h^{-1}~\msun$ from \citet{Yang09}, while the black dashed and dotted lines denote the  satellite counts for host halos with $M_{200} = 10^{13-14}~h^{-1}~\msun$ and satellites at projected distances of $< 400~{\rm kpc}$ within the TNG100 and TNG300 simulations, respectively.
 We find excellent agreement between our inferred satellite counts and those based on simulations and shallower spectroscopic samples.}
 \label{fig:Nsat}
\end{figure}

Selecting satellites within $400~{\rm kpc}$, we find excellent agreement between our inferred satellite stellar mass function and that measured for the IllustrisTNG hosts. As shown in Figure~\ref{fig:Nsat}, our integrated satellite counts are very tightly bracketed by the corresponding predicted counts in the TNG100 and TNG300 simulations, where we select satellites at projected distances of $< 400~{\rm kpc}$ for hosts with $M_{200} = 10^{13-14}~h^{-1}~\msun$. 
In addition, we compare to the observed satellite mass function from \citet{Yang08,Yang09}, based on a sample of spectroscopically-confirmed satellites in $\sim300{,}000$ low-$z$ groups \citep[see also][]{VM2020}.
Overall, our measured satellite mass function is in remarkably good agreement, especially at low masses (or faint magnitudes). While our background-subtraction technique is unable to identify individual satellite galaxies, it is quite robust at indirectly identifying the satellite population, such that its properties may be characterized.


\subsection{Measuring the Satellite Quenched Fraction}
\label{subsec:qf}

As a benchmark for comparison, we measure the quenched fraction as a function of satellite stellar mass for the spectroscopically-confirmed satellites in the \citet{Yang07} group catalog. We limit our sample of host halos to those with $10^{13}~h^{-1}~\msun< \mhalo < 10^{14}~h^{-1}~\msun$, identifying satellites as quenched according to the sSFR cut of $10^{-11}~{\rm yr}^{-1}$ described in \S\ref{subsec:3.1}. 
Unlike our photometric analysis, however, we include groups across the entire SDSS spectroscopic footprint --- i.e.~both within and beyond the Stripe 82 footprint. From this parent population, we then select two subsamples at $z < 0.06$ and at $z < 0.1$. The lower-$z$ ($z < 0.06$) sample includes $\sim1500$ groups with $\sim14{,}000$ satellite galaxies, complete down to a stellar mass of $\sim10^{10}~\msun$. The higher-$z$ sample includes more host systems ($\sim8000$ groups with $\sim40{,}000$ satellites), but only probes down to $\sim10^{10.5}~\msun$. 
In agreement with many previous studies of satellite properties at $z \sim 0$ \citep[e.g.][]{Baldry06, Wetzel13, Woo13, Hirschmann14, Omand14}, we find that the satellite quenched fraction decreases with decreasing satellite stellar mass, such that nearly all satellites are quenched at $>10^{11}~\msun$ with a quenched fraction of $<50\%$ at $\sim10^{10}~\msun$.


\begin{figure*}
 \centering
 \hspace*{-0.001in}
 \includegraphics[width=0.85\textwidth]{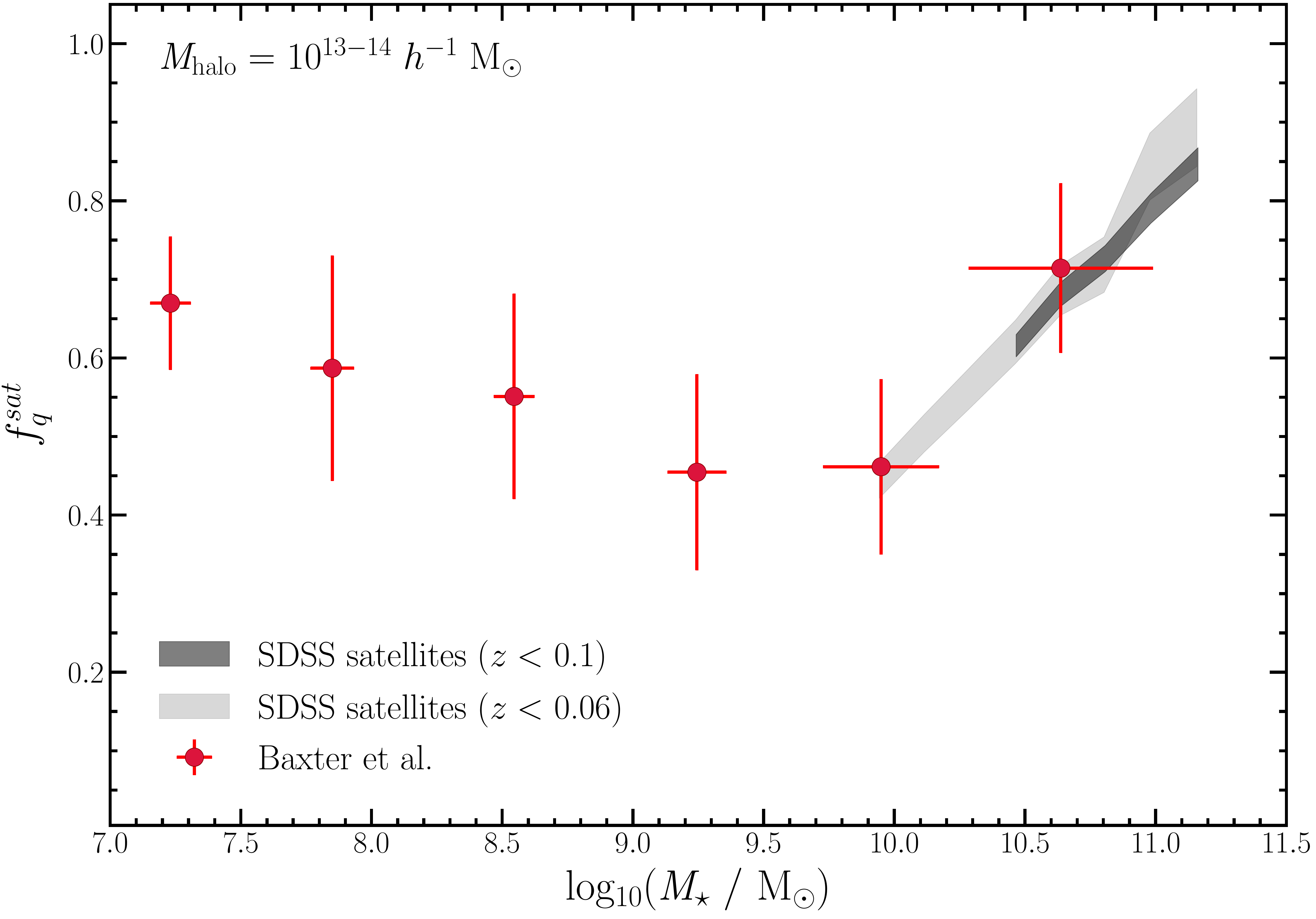}
\caption{The satellite quenched fraction as a function of stellar mass for group environments with $\mhalo = 10^{13-14}~h^{-1}~\msun$. The solid red points represent the median quenched fraction for our statistically-derived satellite population. The vertical error bars correspond to the $1\sigma$ Poisson error in the quenched fraction, while the horizontal error bars denote the standard deviation of the binned stellar masses. The shaded grey (light-grey) band represents the quenched fraction for the spectroscopic members of the \citet{Yang07} groups at $z \lt 0.1$ ($z \lt 0.06$). Our statistically-driven approach using S82 photometry successfully reproduces the satellite quenched fraction results at high masses ($\gt 10^{10}~\msun$), and pushes beyond previous studies to probe satellite quenching down to $10^{7}~\msun$. We find an increase in the quenched fraction at low masses ($\lesssim 10^{9}~\msun$), potentially indicating an increase in the efficiency of quenching in the low-mass regime.}
 \label{fig:fq_v_mstar}
\end{figure*}


In an effort to push measurements of the satellite quenched fraction to lower masses (i.e.~$<10^{10}~\msun$), we use the background subtraction technique described in \S\ref{subsec:bs} as applied to our photometric sample in Stripe 82. Accordingly, we compute the satellite quenched fraction as a function of stellar mass as 
\begin{equation}
f^{sats}_{q}(d_{\rm proj}, \mstar) = \frac{\bar{N}_{\rm sats, q}}{\bar{N}_{\rm sats, q} + \bar{N}_{\rm sats, sf}} \; ,
\label{equation:fq}
\end{equation}
where $\bar{N}_{\rm sats, sf}$ and $\bar{N}_{\rm sats, q}$ are the average number of star-forming and quenched satellites detected at $d_{\rm proj} \lt 400~{\rm kpc}$, respectively. 
As discussed in \S\ref{subsec:bs}, we adopt $400~{\rm kpc}$ as the outer extent of ours groups 
based upon a comparison to comparable halos in the IllustrisTNG simulation suite.
Our resulting satellite quenched fraction, however, remains qualitatively unchanged when integrating satellite counts out to $600~{\rm kpc}$ or $800~{\rm kpc}$.

Figure~\ref{fig:fq_v_mstar} shows the measured satellite quenched fraction as a function of satellite stellar mass using the spectroscopic group membership and our photometric analysis. 
For the stellar mass range at which both approaches overlap (i.e.~$\mstar > 10^{10}~\msun$), we find excellent agreement between the independent measurements. This serves as a strong validation of the background-subtraction technique and our classification model. 

Using the deeper photometry in Stripe 82, we are able to push our measurements of the satellite quenched fraction down to $\sim 10^{7}~\msun$, probing satellite quenching in group environments across four orders of magnitude in satellite stellar mass. 
In contrast to measurements in the high-mass regime ($>10^{10}~\msun$), we find that the satellite quenched fraction in $\mhalo \sim 10^{13-14}~h^{-1}~\msun$ groups increases below satellite stellar masses of $\sim10^{9}~\msun$. 
This transition in the quenched fraction suggests a change in the quenching efficiency (and possibly dominant quenching mechanism), such that the suppression of star formation in low-mass satellites is increasingly efficient at $\mstar \lesssim 10^{9}~\msun$.


\section{Summary and Discussion}
\label{sec:summmary_and_discussion}

We have utilized a combination of supervised machine learning and statistical background subtraction to measure the satellite quenched fraction in group environments across four orders of magnitude in satellite stellar mass ranging from $\mstar \sim 10^{7-11}~\msun$. 
Our analysis utilizes a neural network classifier trained on a spectroscopic training set to label galaxies in the co-added Stripe 82 photometric catalog as either star forming or quenched based solely on their $g-r$, $g-i$, and $r-i$ colors. The results from this procedure were subsequently used to statistically identify the quenched and star-forming satellite populations around spectroscopically-confirmed hosts within Stripe 82 with halo masses between $10^{13-14}~h^{-1}~\msun$. 
The main results from this analysis are as follows: 
\begin{enumerate}[leftmargin=0.25cm]

\item Using our photometric approach, we successfully reproduce the measured satellite quenched fraction at $\mstar \gtrsim 10^{10}~\msun$, as derived from spectroscopic studies in the local Universe. We find that the satellite quenched fraction increases with increasing satellite mass at $\mstar \gtrsim 10^{10}~\msun$. \\

\item We measure the satellite quenched fraction down to $\mstar \sim 10^{7}~\msun$, pushing measurements of satellite quenching in $\sim10^{13-14}~h^{-1}~\msun$ halos to a new regime that is not readily probed outside of the Local Group. \\

\item We find that the satellite quenched fraction increases towards lower satellite masses below  $\sim 10^{9}~\msun$. \\

\item The increase in satellite quenching at low masses potentially indicates a change in the dominant quenching mechanism at $\sim 10^{9}~\msun$, where ram-pressure stripping begins to become increasingly effective (see discussion that follows).

\end{enumerate}

Given that low-mass field galaxies are almost entirely star forming as a population, the increase in the satellite quenched fraction at $<10^{9}~\msun$ can be interpreted as a corresponding increase in the satellite quenching efficiency within $10^{13-14}~h^{-1}~\msun$ halos.
This increase is similar to that observed in the Local Group, where there is an apparent transition in the dominant quenching mechanism at $\sim10^{8}~\msun$ with lower-mass satellites quenched more efficiently following infall. Both hydrodynamic simulations and analytical modeling of the satellite population find that ram-pressure stripping is much more efficient below $10^{8}~\msun$ within Milky Way-like galaxies \citep{Mayer07, Fillingham16, Simpson18, Akins20}, while more massive satellites are primarily quenched via starvation \citep{Fillingham15}.  
Given that our host sample is more massive ($\mhalo = 10^{13-14}~\msun$) relative to Milky Way-like halos, it is expected that an increase in infall velocities and the density of the circumgalactic medium would cause this transition mass to increase, such that starvation is the primary driver of satellite quenching above $\sim10^{9.5}~\msun$ and ram-pressure stripping becoming increasingly important in the low-mass regime. 
A more detailed study of the potential quenching mechanisms at play requires further analysis of the timescales on which the observed satellites are quenched following infall to the host halos. In future work (Baxter et al.~in prep), we aim to bridge this gap by combining the measured satellite quenched fractions from this work with the accretion and orbital histories determined using high-resolution cosmological simulations, to estimate the typical quenching timescale as a function of satellite mass.

The satellite quenched fractions that we obtain at low-masses ($\mstar \lt 10^{9}~\msun$) are generally lower than what have been reported in studies of dwarf galaxies in more massive nearby clusters. For example, \cite{Weinmann11} studied the satellite galaxy population in the nearby Virgo ($\mhalo \sim 1.4–4 \times 10^{14}~\msun$), Coma ($\mhalo \sim 1.3 \times 10^{15}~\msun$), and Perseus ($\mhalo \sim 6.7 \times 10^{14}~\msun$) clusters, finding red fractions between $70-80\%$ at stellar masses of $\sim10^{8-10}~\msun$ (see also \citealt{Boselli16}). 
At slightly higher redshift ($z \sim 0.2$), analysis of the satellite population in Abell 209 ($\mhalo \sim 10^{15}~\msun$) by \citet{Annunziatella16} also finds an elevated quenched fraction relative to our results in less massive halos. Interestingly, while the study of \citet{Annunziatella16} only probes down to $\sim10^{8.6}~\msun$ in satellite stellar mass, the results show a quenched fraction that decreases from near unity ($\sim95\%$) at $\mstar \sim 10^{10.5}~\msun$ to $\sim75\%$ at $\mstar \sim 10^{9}~\msun$ (see also Sarrouh et al.~in prep). 
Naively, if there is a transition in the dominant quenching mechanism (or efficiency) in these massive clusters similar to that found in the Local Group and our group sample, we would expect the transition scale to occur at higher satellite masses (e.g.~$\gtrsim10^{9.5}~\msun$) as ram-pressure stripping (and other cluster-specific processes) should be increasingly effective in hosts with $\mhalo \sim10^{15}~\msun$. 
Extrapolations of the mass functions from \citet{Annunziatella16}, however, do not support this picture. 

Finally, we report satellite quenched fractions in the low-mass regime ($<10^{8}~\msun$) that are potentially lower than expected when compared to studies of satellite quenching in the Local Group, where $\sim90\%$ of satellites with $\mstar < 10^{8}~\msun$ are passive. 
As discussed above, environmental quenching mechanisms are expected to be more efficient in our more-massive host halos relative to the Local Group. 
Of course, our results are based on a study of $\sim100$ groups, whereas studies of the Local Group satellites sample only two host halos. 
While observations of the nearby M81 group yield a satellite quenched fraction comparable to that measured for the Local Group \citep{Kaisin13, Karachentsev13}, various studies also indicate that the Local Group satellites may be outliers relative to the cosmic mean \citep[e.g.][]{BK10, Busha11, Tollerud11, Ibata13, Pawlowski20}. 
Moreover, recent results from the Satellites Around Galactic Analogs (SAGA) Survey \citep{Geha17, Mao20} find lower satellite quenched fractions ($\sim20\%$) around hosts with halo masses comparable to those of the Milky Way and M31. 
We contend that the application of the methodology presented in this work to Milky Way-like hosts is an intriguing way to better place the Local Group into a cosmological context and constrain the quenching of satellites around hosts with $\mhalo \sim 10^{12}~\msun$.

\section*{acknowledgements} 
DCB thanks the LSSTC Data Science Fellowship Program, which is funded by LSSTC, NSF Cybertraining Grant $\#$1829740, the Brinson Foundation, and the Moore Foundation; participation in the program has greatly benefited this work.
MCC thanks Benedetta Vulcani for helpful discussions that improved this work.
We also thank the anonymous referee for helping to improve the clarity of this work.

This work was supported in part by NSF grants AST-1815475 and AST-1518257. Additional support was provided by NASA through grant AR-14289 from the Space Telescope Science Institute, which is operated by the Association of Universities for Research in Astronomy, Inc., under NASA contract NAS 5-26555. 
This research made extensive use of {\texttt{Astropy}},
a community-developed core Python package for Astronomy
\citep{Astropy13, Astropy18}.
Additionally, the Python packages {\texttt{NumPy}} \citep{numpy},
{\texttt{iPython}} \citep{iPython}, {\texttt{SciPy}} \citep{SciPy20}, {\texttt{Scikit-learn}} \citep{scikit-learn}, {\texttt{Keras}} \citep{keras}, and
{\texttt{matplotlib}} \citep{matplotlib} were utilized for our data
analysis and presentation. 

\section*{Data availability} 
Data sharing is not applicable to this article as no new data were created or analyzed in this study.  

\bibliography{citations}

\label{lastpage}
\end{document}